\documentclass[prl,twocolumn]{revtex4}
\usepackage{hyperref}
\usepackage{latexsym}
\def\d{{\mathrm{d}}}

\newcommand{\betrag}[1]{\left\vert#1\right\vert}

\def\d{{\mathrm{d}}}

\def\det{{\mathrm{det}}}
\def\tr{{\mathrm{tr}}}

\def\Box{\kern0.5pt{\lower0.1pt\vbox{\hrule height.5pt
      width 6.8pt \hbox{\vrule width.5pt height6pt
        \kern6pt \vrule width.3pt} \hrule height.3pt
      width 6.8pt} }\kern1.5pt}

\begin{document}
\title{{Naturalness in an emergent analogue spacetime}}
\date{23 December 2005; revised 11 February 2006; \LaTeX-ed \today}
\author{Stefano Liberati }
\email{liberati@sissa.it}
\affiliation{International School for Advanced Studies and INFN,\\ 
Via Beirut 2-4, 34014 Trieste, Italy,}
\author{Matt Visser}
\email{matt.visser@mcs.vuw.ac.nz}
\author{Silke Weinfurtner}
\email{silke.weinfurtner@mcs.vuw.ac.nz}
\affiliation{
School of Mathematics, Statistics,  and Computer Science,\\
Victoria University of Wellington, PO Box 600, Wellington, 
New Zealand}

\begin{abstract}
Effective field theories (EFTs) have been widely used as a framework in order to place constraints on the Planck suppressed Lorentz violations predicted by various models of quantum gravity. There are however technical problems in the EFT framework when it comes to ensuring that small Lorentz violations remain small --- this is the  essence of the  ``naturalness'' problem. Herein we present an ``emergent'' space-time model, based on the ``analogue gravity'' programme, by investigating a specific condensed-matter system  that is in principle capable of simulating 
the salient features of an EFT framework with Lorentz violations.
Specifically, we consider the class of two-component BECs subject to laser-induced transitions between the components,  and we show that this model is an example for Lorentz invariance violation due to ultraviolet physics.   Furthermore our model explicitly avoids the  ``naturalness problem'', and makes specific suggestions regarding how to construct a  physically reasonable quantum gravity phenomenology.
\end{abstract}
\maketitle

\emph{Introduction:}
The purpose of quantum gravity phenomenology (QGP) is to analyze the physical consequences arising from various models of quantum gravity (QG). One hope for obtaining an experimental grasp on QG is the generic prediction arising in many (but not all) quantum gravity models that ultraviolet physics at or near the Planck scale, $M_{\mathrm{Planck}} = 1.2 \times 10^{19} \; \mathrm{GeV/c^2}$, (or in some models the string scale), typically induces violations of  Lorentz invariance (LI) at lower scales~\cite{LIV, jlm-ann}.  Interestingly most investigations, even if they arise from quite different 
fundamental physics, seem to converge on the prediction that the breakdown of Lorentz invariance (LI) can generically become manifest in the form of modified dispersion relations exhibiting extra energy-dependent  or momentum-dependent terms, apart from the usual quadratic one occurring  in the Lorentz invariant dispersion relation. In particular one most often considers Lorentz invariance violations (LIV) in the boost subgroup, leading to an expansion of the  dispersion relation in momentum-dependent terms,
\begin{equation}
 \label{disp}
\omega^2 = \omega_0^2 + \left(1 + \eta_{2} \right) \, c^2 \; k^2 + \eta_{4} \, \left(\frac{\hbar}{M_{\mathrm{LIV}}} \right)^2  \; k^4 + \dots \; ,
\end{equation}
where the coefficients $\eta_{n}$ are dimensionless (and possibly dependent on the particle species considered), and we have restricted our expansion to $CPT$ invariant terms (otherwise one would also get odd powers in $k$). The particular inertial frame for these dispersion relations is generally specified to be the frame set by Cosmological Microwave Background (CMB), and $M_{\mathrm{LIV}}$ is the scale of Lorentz symmetry breaking which furthermore is generally assumed to be of the order of $M_{\mathrm{Planck}}$.

Although several alternative scenarios have been considered in the literature  in order to justify modified kinematics of the kind of Eq.~(\ref{disp}), so far the most commonly explored avenue is an effective field theory (EFT) approach. 
In the present article we wish to focus on the class of non-renormalizable EFTs with Lorentz violations associated to dispersion relations like Eq.~(\ref{disp}). Relaxing our CPT invariance condition this class would include the model developed in~\cite{MP}, and subsequently studied by several authors, where an extension of QED including only mass dimension five Lorentz-violating operators was considered. (That ansatz leads to order $k^3$ LI and CPT violating terms in the dispersion relation.)  Very accurate constraints have been obtained for this model using a combination of  experiments and observations (mainly in high energy astrophysics). See \emph{e.g.}\/~\cite{jlm-ann}. 

In spite of the remarkable success of this framework as a ``test theory'', it is interesting to note that there are still significant open issues concerning its theoretical foundations. 
This is often referred to as the {\em naturalness problem} and can be expressed in the following way. Looking back at our ansatz  (\ref{disp}) we can see that the lowest-order correction, proportional to $\eta_{2}$, is not explicitly Planck suppressed.  This implies that such term would always be dominant with respect to the higher order ones and grossly incompatible with observations (given that we have very good constraints on the universality of the speed of light for different elementary particles).  Following the observational leads it has been therefore often assumed either that some symmetry (other than Lorentz invariance) enforces the $\eta_2$ coefficients to be exactly zero, or that the presence of  some other characteristic EFT mass scale $\mu\ll M_{\rm Planck}$ (\emph{e.g.}, some particle physics mass scale) associated with the Lorentz symmetry breaking might enter in the lowest order dimensionless coefficient $\eta_{2}$ --- which will be then generically suppressed by appropriate ratios of this characteristic mass to the Planck mass: $\eta_2\propto (\mu/M_{\rm Pl})^\sigma$ where $\sigma\geq 1$ is some positive power (often taken as one or two). If this is the case then one has two distinct regimes: For low momenta $p/(M_{\rm Pl}c) \ll (\mu/M_{\rm Pl})^\sigma$ the lower-order (quadratic in the momentum) deviations in~(\ref{disp}) will dominate over the higher-order ones, while at high energies $p/(M_{\rm Pl}c) \gg (\mu/M_{\rm Pl})^\sigma$ the higher order terms will be dominant.

The naturalness problem arises because such a scenario is not well justified within an EFT framework; in other words there is no natural suppression of the low-order modifications in these models. In fact we implicitly assumed that there are no extra Planck suppressions hidden in the dimensionless coefficients $\eta_n$ with $n>2$. EFT cannot justify why \emph{only} the dimensionless coefficients of the $n\leq 2$ terms should be suppressed by powers of the small ratio $\mu/M_{\rm Pl}$.  Even worse, renormalization group arguments seem to imply that a similar mass ratio, $\mu/M_{\rm Pl}$ would implicitly be present also in \emph{all} the dimensionless $n>2$ coefficients --- hence suppressing them even further, to the point of complete undetectability.  Furthermore it is easy to show~\cite{Collins} that, without some protecting symmetry, it is generic that radiative corrections due to particle interactions in an  EFT with only Lorentz violations of order $n>2$ in (\ref{disp}) for the free particles, will generate $n= 2$ Lorentz violating terms in the dispersion relation, which will then be dominant.  Observational evidence~\cite{Coleman-Glashow,LIV} suggests that for a variety of standard model particles $|\eta_2|\lesssim 10^{-21}$. Naturalness in EFT would then imply that the higher order terms are at least as suppressed as this, and hence beyond observational reach.

In order to contribute to this debate, we have chosen a rather unconventional path: We have investigated a condensed matter analog model (AM) of an emergent spacetime~\cite{AM},
that reproduces the salient features of the the non-renormalizable EFT with LIV adopted in quantum gravity phenomenology studies. In particular we looked for a condensed matter system characterized by 1) relativistic kinematics for the low-energy quasi-particles  2) presence of Lorentz violation at high energies due to the underling microscopic structure 3) coexistence on the same background of more than one quasi-particle species (in order to be able to see the presence of deviations from LI at order $k^2$).

Standard Bose--Einstein condensates (BEC) are in this sense interesting systems as they fulfill  the first two of the above requirements (see e.g.\/~\cite{Garay,BEC1}): their excitations can be described at low energy as relativistic phonons propagating on a geometrical background, and at higher order their dispersion relations show modifications of order $k^4$ (this is the so called Bogoliubov dispersion relation, which has been experimentally confirmed~\cite{Vogels:2002qd}). Unfortunately in a single BEC system there is only one species of phonon, and hence it is impossible to address the question of naturalness.
For this reason we chose to investigate the energy dependent behavior of quasi-particles in a 2-component Bose--Einstein condensate. For a specific choice of parameters such a system allows both a massless and a massive quasi-particle (see \cite{2BEC}), which share the same relativistic causal structure in the low energy limit. It is then natural to investigate the expected violation of Lorentz invariance as the high energy regime of the theory is probed. 
(For a detailed discussion, see~\cite{QGPAM}.)


\emph{Sound waves in 2-component BECs:}
The basis for our model is an ultra-cold dilute atomic gas of $N$ bosons in two single-particle states $|A\rangle$ and $|B\rangle$.  For example one could consider a two-component condensate of ${}^{87}$Rb atoms in different hyperfine levels (see e.g.~\cite{Myatt:1997}.)  
The two states have slightly different energies, which permits us, from a theoretical point of view, to keep $m_{A} \neq m_{B}$, even if in experimentally realizable situations $m_A\approx m_B$.  There are three atom-atom coupling constants, $U_{AA}$, $U_{BB}$, and $U_{AB}$, and for our purposes it is essential to include an additional coupling $\lambda$ that drives transitions between the two single-particle states. 
For temperatures at or below the critical BEC temperature, almost all atoms occupy the respective ground states $|A\rangle,|B\rangle$ and it is hence meaningful to adopt the mean-field description for these modes. Ignoring back reaction effects of the quantum fluctuations one then obtains a pair of coupled Gross--Pitaevskii equations (GPE)
\begin{eqnarray}  \label{2GPE} 
 i \, \hbar \, \partial_{t} \Psi_{i} &=& \bigg[
   -\frac{\hbar^2}{2\,m_{i}} \nabla^2 + V_{i}-\mu_{i} 
 \nonumber
 \\
 &&
   + U_{ii}
   \, \betrag{\Psi_{i}}^2 + U_{ij} \betrag{\Psi_{j}}^2
   \bigg] \Psi_{i} 
    + \lambda \, \Psi_{j} \, , 
\end{eqnarray}
where  $(i,j)\rightarrow (A,B)$ or  $(i,j)\rightarrow (B,A)$ and $\Psi_i$ is the classical wave function of the condensate $\langle \Psi\rangle$. 
Now consider small perturbations (sound waves) in the condensate cloud.  The excitation spectrum is obtained by linearizing around some background, and after a straightforward analysis leads to a differential equation for the (rescaled) perturbations in the phases, $\tilde{\theta}=\Xi^{-1/2}\bar\theta=\Xi^{-1/2}[\theta_{A1},\theta_{B1}]^T$, where $\Xi$ is a $2\times2$ matrix constructed from the atomic couplings $U_{ij}$ \cite{QGPAM}.

\emph{Hydrodynamic limit:} If one ignores the effect of the quantum potential then one obtains 
\begin{eqnarray} \label{phaseequation2}
\partial_{t}^2\tilde{\theta} &=&
 - \partial_{t} \left(\mathbf{I} \; \vec v_0 \cdot \nabla \tilde{\theta} \right) 
 - \nabla    \cdot   \left(\vec v_0 \; \mathbf{I} \; \dot{\tilde{\theta}} \right)  
 \nonumber
 \\
 &&
 + \nabla \cdot \left[ \left(C_0^2 - \vec v_0 \; \mathbf{I} \; \vec v_0  \right) \nabla \tilde{\theta} \right] 
 + \Omega^2 \; \tilde{\theta},
\end{eqnarray}
where $C_{0}^2$ and  $\Omega^2$ are $2\times2$ symmetric matrices constructed from the parameters appearing in the GPE, and $\vec v_0$ is the common flow velocity of the background condensates.  If  $ [C_{0}^2, \; \Omega^2] = 0$ then decomposition onto the eigenstates of the system results in a pair of independent ``curved spacetime'' Klein--Gordon equations
\begin{equation} \label{KGE}
\frac{1}{\sqrt{-g_{\mathrm{I/II}}}}
\partial_{a} \left\{   \sqrt{-g_{\mathrm{I/II}}} \; (g_{\mathrm{I/II}})^{ab} \; 
\partial_{b} \tilde{\theta}_{\mathrm{I/II}} \right\} + 
\omega_{\mathrm{I/II}}^2 \;
\tilde{\theta}_{\mathrm{I/II}} = 0,
\end{equation}
where the ``acoustic metrics'' are given by
\begin{equation} \label{metric}
(g_{\mathrm{I/II}})_{ab}\propto
\left[
\begin{array}{ccc}
-\left( c_{\mathrm{I/II}}^2-v_0^2 \right)       &|& -\vec{v_0}^{\,T} \\
\hline
-\vec{v_0}  &|& \mathbf{I}_{d\times d}
\end{array}
\right] \, ,
\end{equation}
and where the overall conformal factor depends on the spatial dimension $d$. 
The metric components depend only on the background velocity $\vec{v}_{0}$, the background densities $\rho_{0i}$,  and the speeds of sound $c^2_{I/II}$ for the two eigenmodes.  These are given by the eigenvalues of the matrix $C_0^2$:
\begin{equation}
\label{e:csq-Xi}
c_{\mathrm{I/II}}^2 = 
\frac{\tr[C_0^2] \pm \sqrt{\tr[C_0^2]^2 - 4 \, \det[C_0^2]}}{2} \,.
\end{equation}
The speed of sound in the AM takes on the role of the speed of light. 
The matrix $\Omega^2$ can be shown to have zero determinant, and so the eigenfrequencies of the two phonon modes are: $\omega_{\mathrm{I}}^2 = 0$ and $\omega_{\mathrm{II}}^2 = \tr[\Omega^2]$.
The masses of the modes are then defined as
$
m_{\mathrm{I/II}}^2 = \hbar^2 \omega_{\mathrm{I/II}}^2/c_{\mathrm{I/II}}^4 
$
and thus the AM corresponds to one massless particle $m_{\mathrm{I}}=0$ and one massive particle.
They both ``experience'' the \emph{same} space-time if the sound speeds are equal, which requires $\tr[C_0^2]^2 - 4 \det[C_0^2]=0$, \emph{i.e.}~$c_I=c_{II}=c_0=\tr[C^2_0]/2$. Hence we now have an AM representing both massive and massless particles, propagating on the same background at low energies. Let us now extend the analysis to high-energies and explore the structure of the corresponding LIV.


\emph{QGP beyond the hydrodynamic limit:}
Starting from the GPE, we now linearize around a uniform condensate and set the background velocity to zero, $\vec{v}_0 =\vec{0}$, but retain the quantum pressure term. The equation for the phase perturbations in momentum space is now \cite{QGPAM}
\begin{eqnarray}
\omega^2 {\tilde{\theta}}  &=& 
\left\{
\sqrt{\Xi+X\; k^2} \;\; [D\; k^2+\Lambda]\;\; \sqrt{\Xi+X\;k^2} 
\right\}\;  \tilde{\theta} 
\nonumber
\\
&=& H(k^2) \; \tilde{\theta} \, ,
   \label{eq:new-disp-rel}
\end{eqnarray}
where $\Xi$, $X$, $D$, and $\Lambda$ are additional $2\times2$ symmetric matrices constructed from the parameters appearing in the GPE.
The perturbation spectrum obeys the generalized Fresnel equation:
\begin{equation}
\det\{ \omega^2 \;\mathbf{I} - H(k^2) \} =0 \, ,
\end{equation}
and the dispersion relations for the phonon modes are
\begin{equation}
\omega_{\mathrm{I/II}}^2 = { \hbox{tr}[H(k^2)] \pm \sqrt{
    \hbox{tr}[H(k^2)]^2 - 4\;  \det[H(k^2)] }\over 2}.
\label{eq:tot-disp-rel}
\end{equation}
A Taylor-series expansion gives
\begin{eqnarray} \label{Taylor} 
\omega_{\mathrm{I/II}}^2 &=& 
\left. \omega^2_{\mathrm{I/II}} \right\vert_{k \rightarrow 0}  
+ \left. \frac{\d \omega_{\mathrm{I/II}}^2}{\d k^2} \right\vert_{k \rightarrow 0} \!\!\! k^2
+ \left. \frac{1}{2} \, \frac{\d^2 \omega_{\mathrm{I/II}}^2}{\d \left(k^2\right)^2} 
\right\vert_{k \rightarrow 0} \!\!\! \left(k^2\right)^2
\nonumber
\\
&&
+ \mathcal{O}\left[( k^2)^3 \right] \, ,
\end{eqnarray}
so these two dispersion relations are in the desired form of Eq (\ref{disp}). Note $\omega_{\mathrm{I/II}}^2 = \omega_{\mathrm{I/II}}^2(k^2)$ only permits even powers in $k$, as the dispersion relation is invariant under CPT. This is by no means a surprising result, because the underlying GPE (\ref{2GPE}) is also invariant under CPT.

It is now useful to define the symmetric matrices $C^2 = C_0^2 + \Delta C^2$ and $\Delta C^2 = X^{1/2}\Lambda X^{1/2}$
which describe how the speed of sound is modified, and also to define  $Y^2 =  2 X^{1/2} \Xi^{-1} X^{1/2}$ and $Z^2 =  2 X^{1/2} D X^{1/2}$. All three of $\Delta C^2$, $Y^2$, and $Z^2$ are explicitly suppressed by powers of the mass of the fundamental constituents that condense to form the BEC.
Note that all the relevant matrices have been carefully symmetrized, and note the important distinction between $C_0^2$ and $C^2$.
Now define $c^2 = {1\over2}\tr[C^2]$,
which approaches the speed of sound $c^2 \rightarrow c_0^2$, in the hydrodynamic limit.
The second and fourth order coefficients in the dispersion relations (\ref{Taylor}) are~\cite{QGPAM}:
\begin{eqnarray} 
 \label{omega_I}
\left.{\d\omega_{\mathrm{I/II}}^2\over\d k^2}\right|_{k\to0} 
&=& c^2 \left[ 1\pm \left\{
 2 \tr[\Omega^2 C_0^2 ] - \tr[\Omega^2]\;\tr[C^2] \over
 \tr[C^2] \tr[\Omega^2]  \right\}
 \right]
 \nonumber
 \\
 &=& c^2 (1\pm \eta_2) \, ; \\ 
\left.{\d^2\omega_{\mathrm{I/II}}^2\over\d(k^2)^2}\right|_{k\to0} &=&  {\textstyle 1\over \textstyle 2} \Bigg[
\tr[Z^2]  \pm \tr[Z^2] 
\nonumber
\\
&&
\pm 2 \frac{\tr[\Omega^2{C}^2_0]-\tr[\Omega^2]\;\tr[C_0^2]}{\tr[\Omega^2]}\tr[Y^2]
\nonumber
\\
&&
\pm {\tr[C^2]^2- 4\det[C^2_0] \over \tr[\Omega^2] }
 \mp
{\tr[C^2]^2\over\tr[\Omega^2]} \eta_2^2
\Bigg] 
\nonumber
\\
&=& 2 \eta_{4} \left( {\hbar}/{M_{\rm LIV}} \right)^2 \; .  \label{omega_II}
\end{eqnarray}

\emph{Lorentz violations from UV physics:}
In order to obtain LIV purely due to ultraviolet physics, we demand exact Lorentz invariance  in the hydrodynamic limit. In other words, we require all terms in the equations (\ref{omega_I}) and (\ref{omega_II}) which might otherwise survive in the hydrodynamic limit to be set to zero. The constraints we obtain are:
 \begin{eqnarray}
&&C1:\qquad \tr[C^2_0]^2-4\det[C^2_0]=0 \, ;\\
&&C2:\qquad 2\tr[\Omega^2 {C}^2_0]-\tr[\Omega^2]\tr[C^2_0]=0 \, .
\end{eqnarray}
Beyond the hydrodynamic limit, but imposing $C1$ and $C2$, the equations (\ref{omega_I}) and (\ref{omega_II}) simplify to:
\begin{eqnarray}
\left.{\d\omega_{\mathrm{I/II}}^2\over\d k^2}\right|_{k\to0} 
&=& c_0^2 + {1\pm1\over2}\tr[\Delta C^2] \, ,\label{eq:varpi2b}
\\
\label{eq:varpi4b}
\left.{\d^2\omega_{\mathrm{I/II}}^2\over\d(k^2)^2}\right|_{k\to0} 
&=&
{\tr[Z^2]  \pm \tr[Z^2] \over 2}
\nonumber
\\
&&
\pm\tr[C^2_0]\left(-\tr[Y^2]+
{\tr[\Delta C^2] \over \tr[\Omega^2] }\right) \, .  \qquad
\end{eqnarray}
To enforce $C1$ and $C2$ the effective coupling between the hyperfine states has to vanish, $ \tilde{U}_{AB}=0$. This can be done by imposing a particular transition rate $\lambda = -2 \sqrt{\rho_{A0}\;\rho_{B0}} \; U_{AB}$.
In addition, fix the hydrodynamic speed of sound to be
\begin{equation}   \label{eq:c0av}
c_0^2 
= \frac{m_B \rho_{A0} U_{AA} + m_A \rho_{B0} U_{BB} + U_{AB} (\rho_{A0} m_A + \rho_{B0} m_B) }{2 m_A m_B }.
\end{equation}
While one eigenfrequency always remains zero, $\omega_{0,\mathrm{I}}\equiv0$, for the second phonon mode we get
\begin{equation}
\omega_{0,II}^2 = \frac{4 U_{AB} (\rho_{A0} m_B + \rho_{B0} m_A) c_0^2}{ \hbar^2} \,.
\label{eq:om2}
\end{equation}
Thus the AM corresponds to one massless particle $m_{\mathrm{I}}=0$ and one massive particle $m^2_{II}=\hbar^2\omega_{0,II}^2/c^4_0$,
propagating in the acoustic Minkowski space-time in the hydrodynamic limit. This mass is much smaller that any average of the atomic masses if we set $U_{AB}\ll U_{AA}+U_{BB}$.  (Although not relevant to the aim of the present paper, we stress that such a regime is potentially achievable in an experimental setting.) 
For higher wave numbers we obtain LIV in the form of equation (\ref{disp}), and the coefficients $\eta_{2}$ and $\eta_{4}$ for the two modes are: 
\begin{eqnarray} 
 \eta_{2,\mathrm{I/II}}
&\approx&\left(\frac{m_{\mathrm{I/II}}}{M_{\rm LIV}}\right)^2= 
\left( \frac{ \mathrm{quasiparticle\;mass} } 
{\mathrm{effective\;Planck\;scale}} \right)^2 \, ; \qquad
\label{eta2_final}
\\
\eta_{4,\mathrm{I/II}} 
 &\approx& 1;
\label{eta4_final}
\end{eqnarray}
where 
$M_{\mathrm{LIV}}=\sqrt{m_{A} m_{B}}$ is defined as the scale of Lorentz violations --- which is our analogue Planck scale. 


It is quite remarkable that the quadratic coefficients (\ref{eta2_final}) are {\em exactly} of the form postulated in several works on non-renormalizable EFT with LIV (see e.g.~\cite{jlm-ann}). They are indeed the squared ratio of the particle mass to the scale of Lorentz violation.
Moreover we can see from (\ref{eta4_final}) that there is no further suppression --- after having pulled out a factor $(\hbar / M_{\mathrm{LIV}})^2$ --- for the quartic coefficients $\eta_{4,\mathrm{I/II}}$. These coefficients are of order one and generically non-universal, (though if desired they can be forced to be universal by additional and specific fine tuning).

\emph{Discussion:}
The suppression of $\eta_2$, combined with the \emph{non-suppression} of $\eta_4$, is precisely the statement that the ``naturalness problem'' does not arise in the current model. We stress this is not a ``tree level'' result as the dispersion relation was computed directly from the fundamental Hamiltonian and was not derived via any EFT reasoning.
 Moreover avoidance of the naturalness problem is not directly related to the tuning of our system to reproduce SR in the hydrodynamic limit. In fact our conditions for recovering SR at low energies do not \emph{a priori} fix the the $\eta_2$ coefficient,  as its strength after the ``fine tuning" could still be large (even of order one) if the typical mass scale of the massive phonon is not well below the atomic mass scale. Instead the smallness of $\eta_2$ is directly related to the mass-generating mechanism.

The key question is now: Why does our model escape the naive predictions of dominant lowest-dimension Lorentz violations?  (In fact in our model for any $p\gg m_{II}$ the $k^4$ LIV term dominates over the order $k^2$ one.)
We here propose a nice interpretation in terms of  ``emergent symmetry'': Non-zero $\lambda$ \emph{simultaneously} produces a non-zero mass for one of the phonons, \emph{and} a corresponding non-zero LIV at order $k^2$. (Single BEC systems have only $k^4$ LIV as described by the Bogoliubov dispersion relation.) Let us now drive $\lambda\to 0$, but keep the conditions $C1$ and $C2$ valid  at each stage. (This also requires $U_{AB}\to 0$.) One gets an EFT which at low energies describes two non-interacting phonons propagating on a common background. (In fact $\eta_2\to0$ and $c_I=c_{II}=c_0$.) This system possesses a $SO(2)$ symmetry. 
Non-zero laser coupling $\lambda$ softly breaks this $SO(2)$, the mass degeneracy, and  low-energy Lorentz invariance. Such soft Lorentz violation is then characterized (as usual in EFT) by the ratio of the scale of the symmetry breaking $m_{II}$, and that of the scale originating the LIV in first place $M_{\rm LIV}$. We stress that the $SO(2)$ symmetry is an ``emergent symmetry'' as it is not preserved beyond the hydrodynamic limit: the $\eta_4$ coefficients are in general different if $m_A\neq m_B$, so $SO(2)$ is generically broken at high energies.  Nevertheless this is enough for the protection of the {\em lowest}-order LIV operators. 
The lesson to be drawn is that emergent symmetries are sufficient to minimize the amount of Lorentz violation in the lowest-dimension operators of the EFT. 

We  acknowledge useful discussions and comments by David Mattingly, Ted Jacobson, 
and Bei-Lok Hu.


\end{document}